\documentclass[10pt]{article}
\usepackage{graphicx}
\usepackage{verbatim}   
\usepackage{color}      
\usepackage{subfigure}  
\usepackage[abs]{overpic}
\usepackage{float}
\restylefloat{table}
\usepackage{color}

\usepackage{lineno}     

\def\institute{
INFN Gruppo Collegato di Udine, Sezione di Trieste, Udine and ICTP, Trieste \\
Strada Costiera 11, Trieste 34151, Italy}

\def\Title#1{\begin{center} {\Large #1 } \end{center}}
\def\Author#1{\begin{center}{ \sc #1} \end{center}}
\def\Address#1{\begin{center}{ \it #1} \end{center}}

\newenvironment{Abstract}{\begin{quotation}  }{\end{quotation}}
\newenvironment{Presented}{\begin{quotation} \begin{center} 
             PRESENTED AT\end{center}\bigskip 
      \begin{center}\begin{large}}{\end{large}\end{center} \end{quotation}}

\def\ttbar{\ensuremath{t\bar{t}}}
\def\pt{\ensuremath{p_{\mathrm{T}}}} 

\def\ttDM{\ensuremath{t\bar{t}+\chi\bar{\chi}}}
\def\ptmiss{{p}_{\rm{T}}^{\rm{miss}}}

\def\met{{E}_{\rm{T}}^{\rm{miss}}}

\def\beq{\begin{equation}}
\def\eeq#1{\label{#1}\end{equation}}
\def\eeqn{\end{equation}}

\def\beqa{\begin{eqnarray}}
\def\eeqa#1{\label{#1}\end{eqnarray}}
\def\eeqan{\end{eqnarray}}

\let\bar=\overbar

\def\Dslash{\not{\hbox{\kern-4pt $D$}}}
\def\dslash{\not{\hbox{\kern-2pt $\del$}}}

\def\msb{{\bar{\ssstyle M \kern -1pt S}}}

\begin{document}
\begin{titlepage}


\vfill
\Title{Top quarks and exotics at ATLAS and CMS}
\vfill
\Author{ Leonid Serkin \\ \normalsize{on behalf of the ATLAS and CMS Collaborations\footnote{Copyright [2018] CERN for the benefit of the ATLAS and CMS Collaborations. Reproduction of this article or parts of it is allowed as specified in the CC-BY-4.0 license}}}
\Address{\institute}
\vfill

\begin{Abstract}
An overview of recent searches with top quarks in the final state using up to 36~fb$^{-1}$ of $pp$ collision data at $\sqrt{s}$ = 13 TeV collected with the ATLAS and CMS experiments at the LHC is presented. In particular, searches for heavy resonances decaying into third generation quarks, searches for production of vector-like quarks and searches for dark-matter produced in association with top quarks are summarised. No significant excess over the SM prediction is found and 95\% CL exclusion limits are set in a variety of models.
\end{Abstract}

\vfill
\begin{Presented}
$11^\mathrm{th}$ International Workshop on Top Quark Physics\\
Bad Neuenahr, Germany, September 16--21, 2018\\
\vfill
ATL-PHYS-PROC-2018-195
\end{Presented}
\vfill
\end{titlepage}
\def\thefootnote{\fnsymbol{footnote}}
\setcounter{footnote}{0}
%

\normalsize 

\section{Introduction}
With a mass close to the scale of electroweak symmetry breaking, the top quark, besides having a large
coupling to the Standard Model (SM) Higgs boson~\cite{bib:ttH_CMS,bib:ttH_ATLAS}, is predicted to have large couplings to new
particles predicted in many models beyond the Standard Model (BSM). Possible new phenomena (referred to as ``exotics'') may enhance the cross sections over SM predictions through the production of heavy particles in association with a top-quark pair or the production of new gauge bosons massive enough as to decay into a third generation quarks.

\medskip
 
\begin{figure*}[h]
\centering
\begin{overpic}[height=0.14\textwidth]{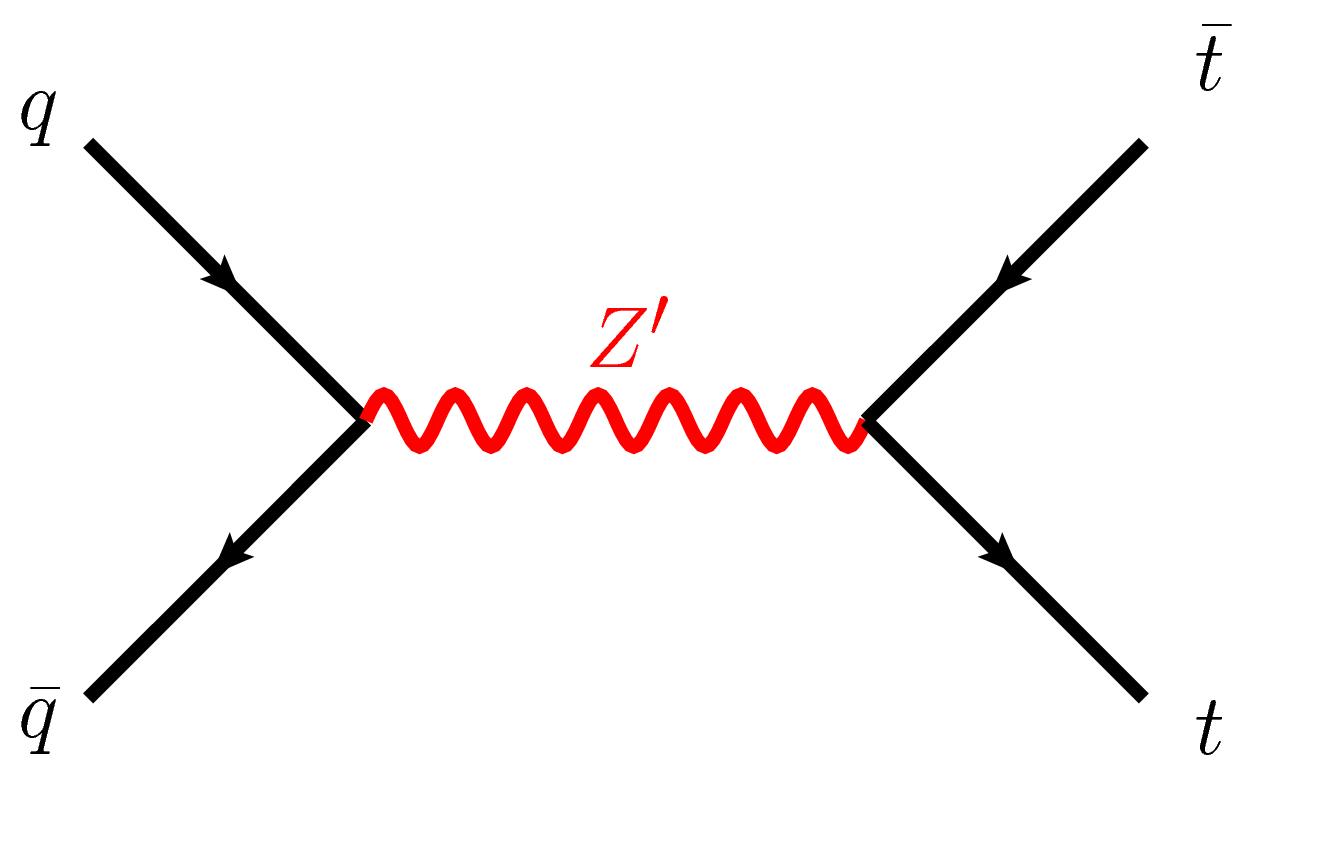}
\put(45,-7){(a)}
\end{overpic}
\begin{overpic}[height=0.14\textwidth]{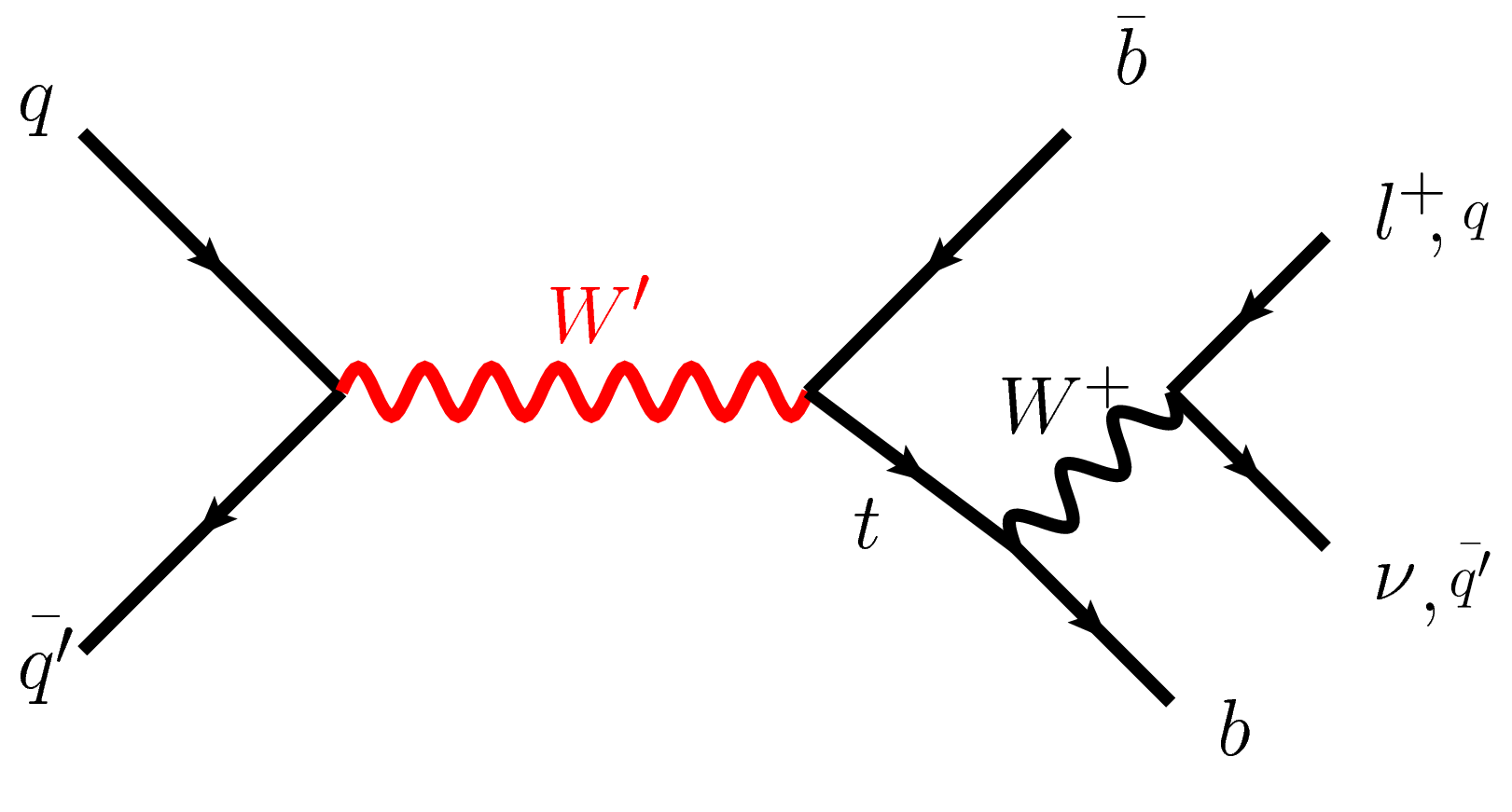}
\put(45,-7){(b)}
\end{overpic}
\qquad
\begin{overpic}[height=0.14\textwidth]{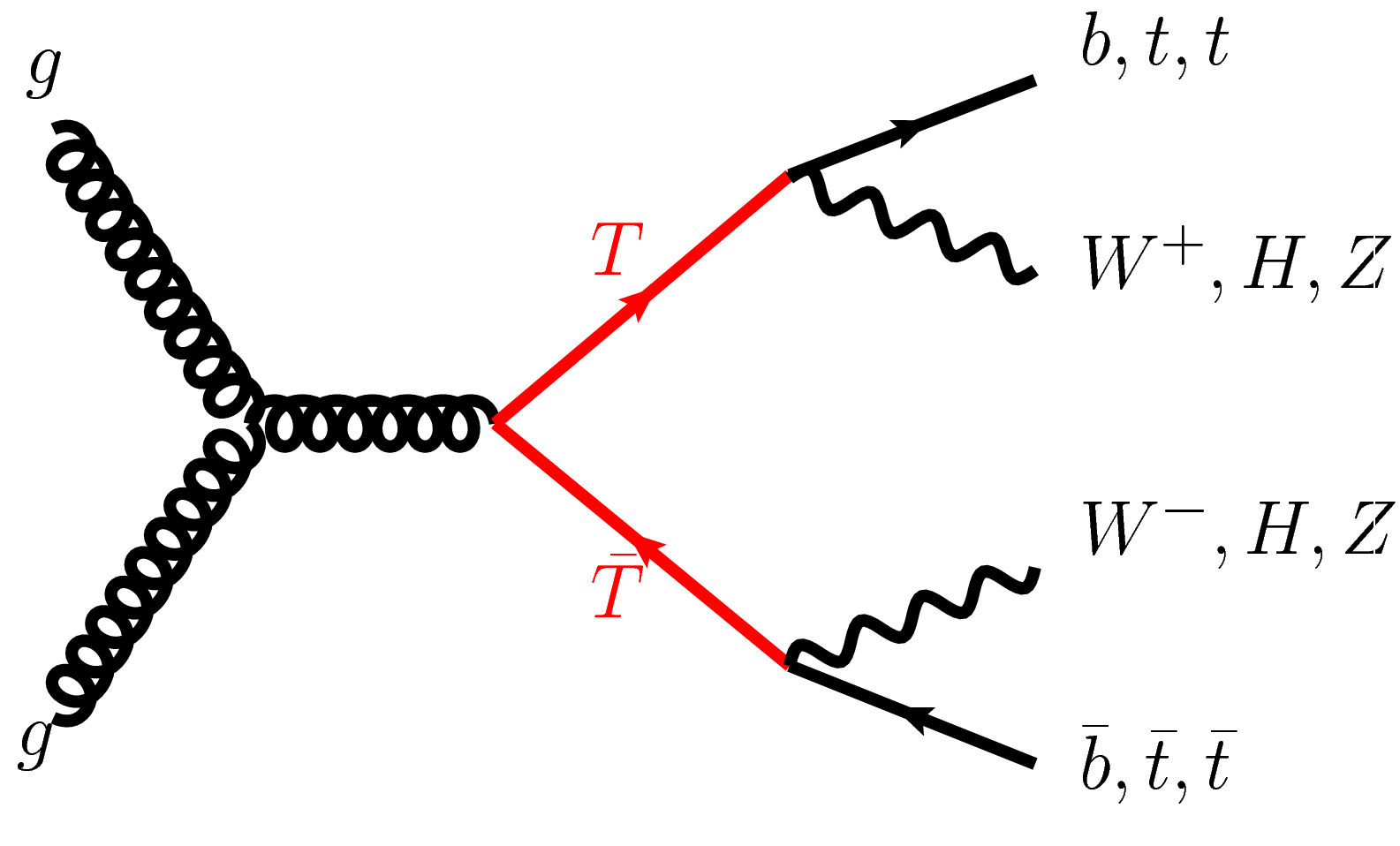}
\put(45,-7){(c)}
\end{overpic}
\begin{overpic}[height=0.14\textwidth]{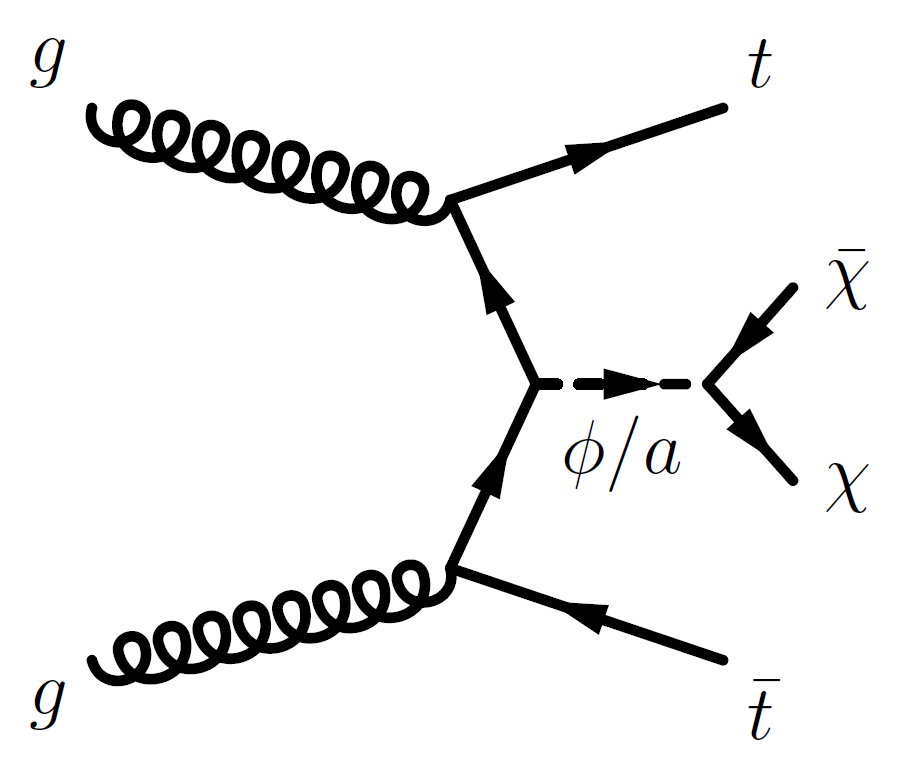}
\put(45,-7){(d)}
\end{overpic}
\caption{
Representative leading-order Feynman diagrams for 
(a) spin-1 $Z^{'}$ boson production, 
(b) $W^{'}$ boson production with decay into $t\bar{b}$ final state, 
(c) vector-like $T$ pair production and its decays, 
(d) colour-neutral spin-0 mediator associated production with top quarks.
}
\label{fig:Plot} 
\end{figure*}

Many SM extensions predict different types of heavy resonances at the TeV scale that decay to $\ttbar$-pairs: color singlet $Z^{'}$ boson depicted in Fig.~\ref{fig:Plot}(a), Kaluza-Klein (KK) excitations of the gluons ($g_{KK}$) or gravitons ($G_{KK}$). Other BSM models introduce new charged vector currents mediated by heavy gauge bosons, usually referred to as $W^{'}$, which are expected to couple more strongly to the third generation of quarks than to the first and second ones. Searches for a heavy $W^{'}$ boson decaying into $t\bar{b}$\footnote{The notation $t\bar{b}$ is used to describe both the $W^{'+} \to t\bar{b}$ and $W^{'-} \to \bar{t}b$ processes.}, as shown in Fig.~\ref{fig:Plot}(b), or heavy resonances decaying to $\ttbar$-pairs, can be classified as either leptonic or hadronic according to the decay products of the $W$-boson originating from the top quark(s). The common signature in both searches is a deviation from the $t\bar{b}$ ($\ttbar$) invariant mass spectrum predicted by the SM.

\medskip

The existence of heavy top quark partners is particularly well motivated to cancel the largest corrections from SM top quark loops. In supersymmetric theories bosonic partners of the top quark serve this purpose, but in several other BSM theories, this role is filled by fermionic top quark partners. These models predict the existence of vector-like quarks (VLQs), color-triplet spin-1/2 fermions whose left- and right-handed chiralities transform in the same way under weak-isospin. VLQs are expected to couple preferentially to third-generation quarks and can have flavour-changing neutral-current decays in addition to charged-current decays.  An up-type vector-like $T$ quark with charge $+2/3$ can decay into $Wb$, $Zt$, or $Ht$, as shown in Fig.~\ref{fig:Plot}(c) for $T\bar{T}$-pair production. Similarly, a down-type quark $B$ with charge $-1/3$ can decay into $Wt$, $Zb$, or $Hb$, giving a very rich phenomenology at the LHC based on the final state signatures.

\medskip
In a large class of models, dark matter (DM) consists of stable, weakly interacting massive particles ($\chi$), which may be pair produced at the LHC via mediators that couple both to DM particles and to SM quarks. A favoured class of models propose a spin-0 mediator with SM Higgs-like Yukawa coupling to quarks, which therefore couples preferentially to the top quark. Consequently, dark matter production can be probed in association with a top quark pair ($\ttDM$) and the dominant mechanism for this production is the $s$-channel production of the mediator via gluon fusion, with the mediator then decaying to a pair of DM particles, as shown in Fig.~\ref{fig:Plot}(d). Usually, the DM particles are assumed to be Dirac fermions, and the mediators are spin-0 particles with scalar ($\phi$) or pseudoscalar ($a$) interactions. The couplings between the mediator and SM quarks are: $g_{qq}= g_{q} y_{q}$, where $y_{q}=\sqrt{2} m_{q} / v$ are the SM Yukawa couplings, $m_{q}$ is the quark mass, and $v=246$ GeV is the Higgs boson field vacuum expectation value. The $g_{q}$ parameter is assumed to be equal to unity for all quarks.

\medskip
In the following, an overview of the latest exotics results with 13 TeV data taken by the ATLAS~\cite{bib:ATLAS_det} and CMS~\cite{bib:CMS_det} experiments at the LHC is presented.

\section{Searches for new heavy particles decaying into $\ttbar$-pairs }
\begin{figure*}[h]
\centering
\begin{overpic}[height=0.27\textwidth]{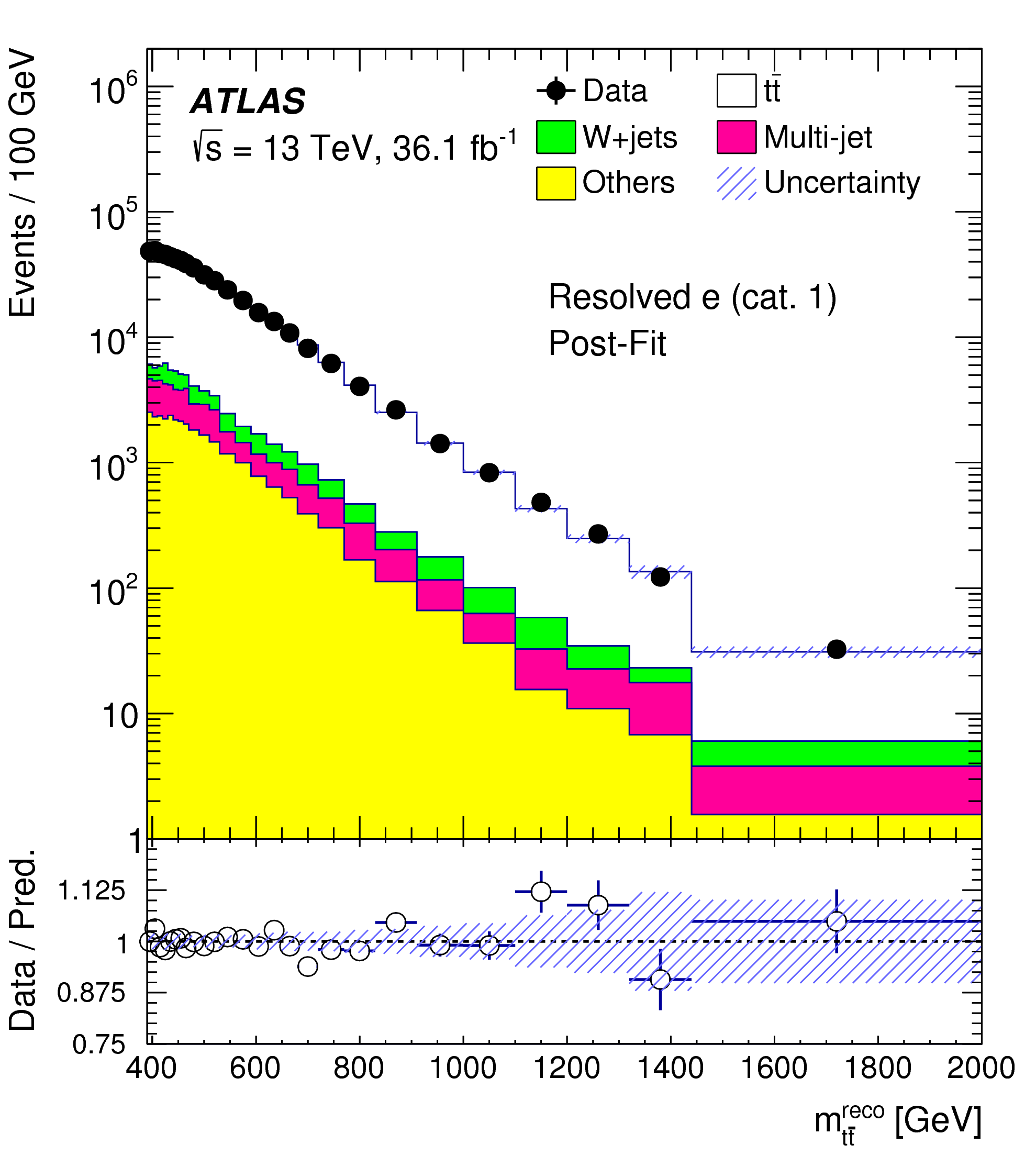}
\put(75,-7){(a)}
\end{overpic}
\qquad
\begin{overpic}[height=0.27\textwidth]{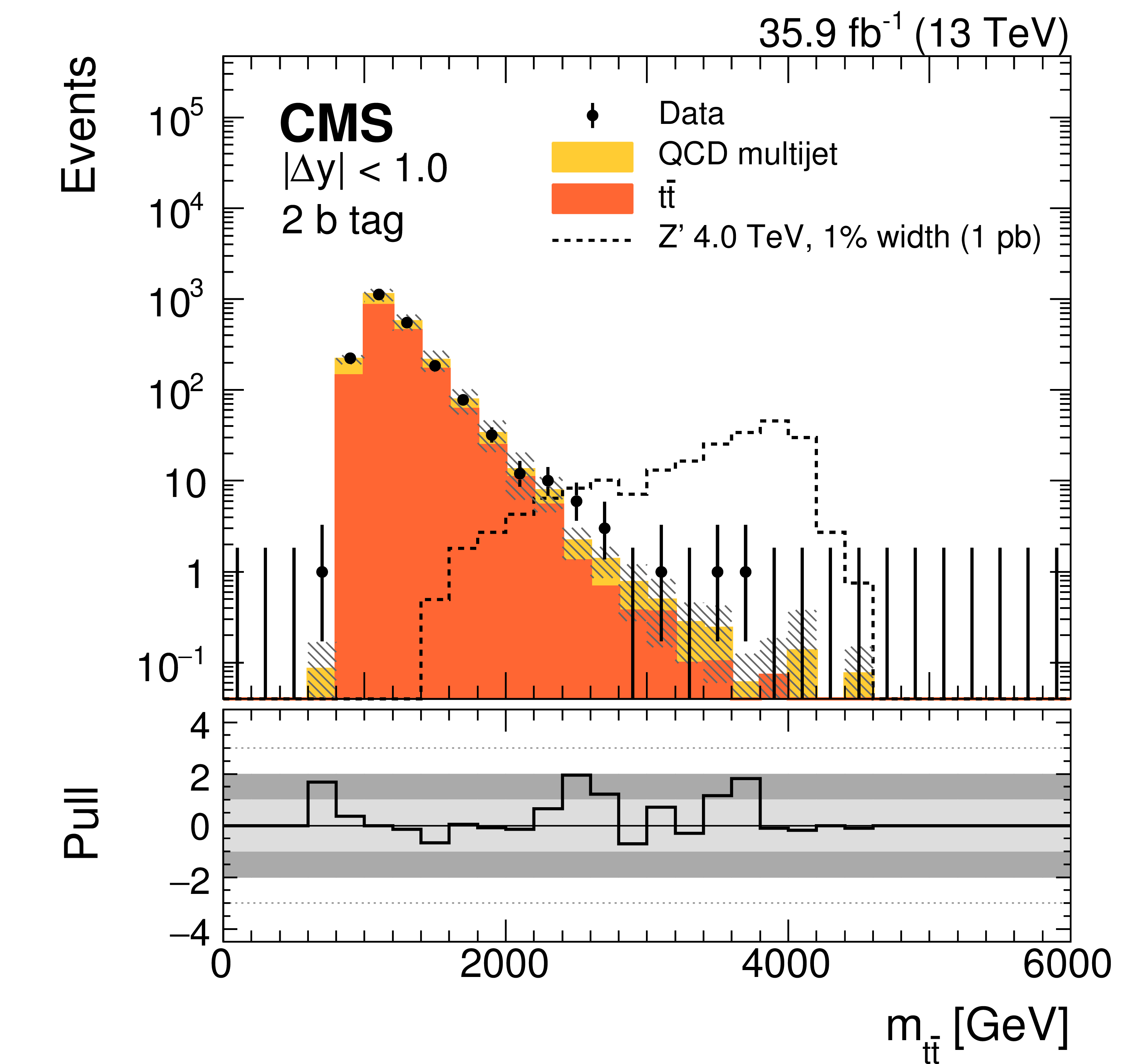}
\put(75,-7){(b)}
\end{overpic}
\qquad
\begin{overpic}[height=0.27\textwidth]{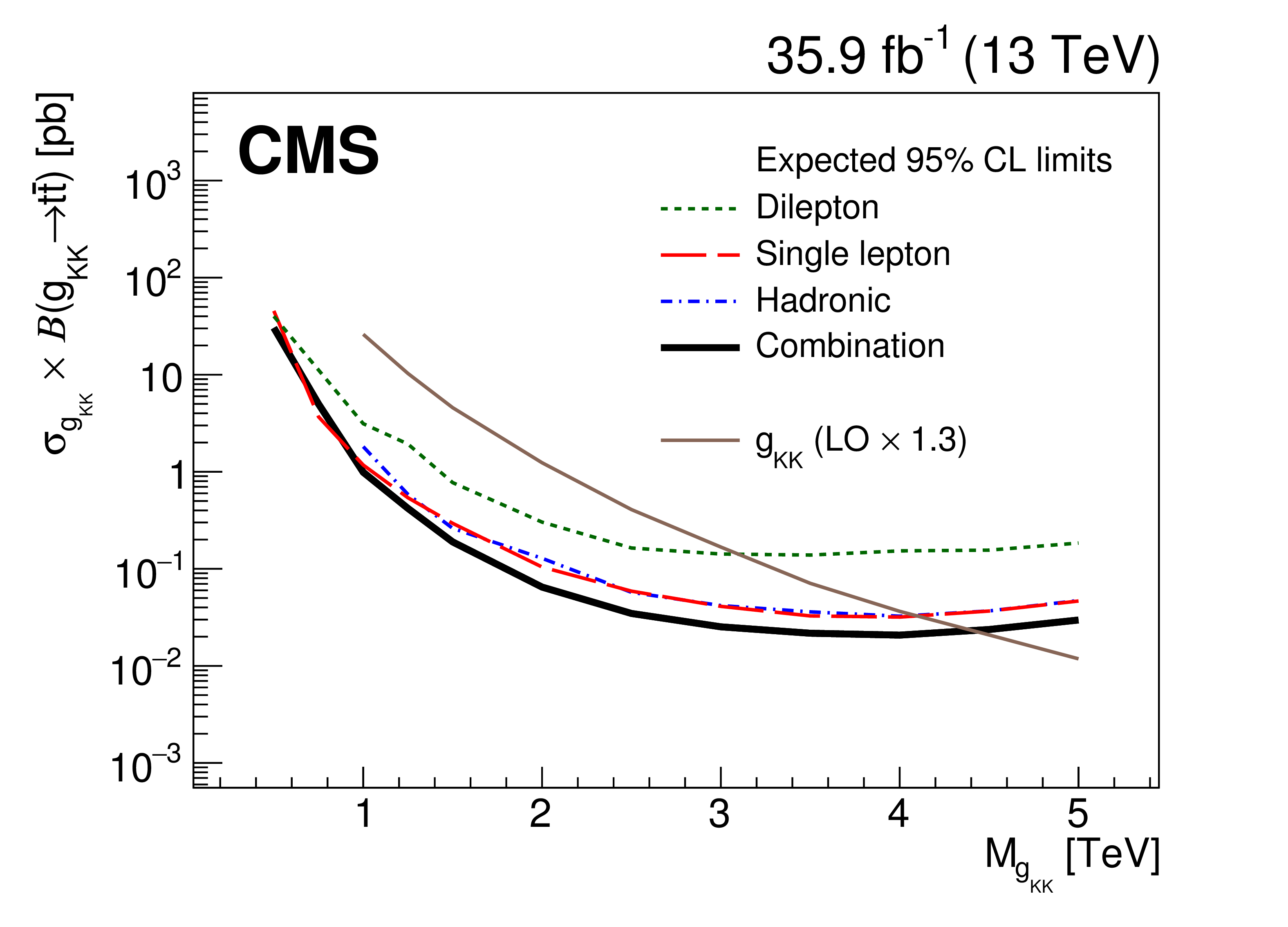}
\put(75,-7){(c)}
\end{overpic}
\caption{(a) Distribution of $m_{\ttbar}^{\rm{reco}}$ distributions for the resolved selection in the ATLAS search~\cite{bib:ATLAS_res}. (b) Distributions of $m_{\ttbar}$ for SRs in the fully hadronic channel in the CMS search~\cite{bib:CMS_res}. (c) Comparison of the sensitivities for each analysis channel in the CMS search contributing to the combination, shown for the $g_{KK}$ signal hypothesis~\cite{bib:CMS_res}.}
\label{fig:Plot1} 
\end{figure*}

A search for heavy particles decaying into top-quark pairs was recently performed by the ATLAS experiment using 36.1~fb$^{-1}$ of $\sqrt{s}$ = 13 TeV $pp$ collision data~\cite{bib:ATLAS_res}, selecting events that are consistent with $\ttbar$ production followed by their subsequent decay into the lepton+jets topology. Based on the hadronic activity, the events are classified into boosted or resolved topologies. The $m_{\ttbar}^{\rm{reco}}$ observable is constructed to approximate the invariant mass of the $\ttbar$ system. For events passing the boosted selection, the value of $m_{\ttbar}^{\rm{reco}}$ is built from the mass of the summed four-momenta of the leptonic (sum of the four-momenta of the charged lepton, the neutrino candidate, and highest-$\pt$ $b$-tagged jet) and hadronic (the four momentum of the highest-$\pt$ top-tagged large-$R$ jet) top candidates. For events passing the resolved selection, a $\chi^{2}$ algorithm is employed to find the best assignment of jets to the leptonic-top candidate and hadronic-top candidate. No excess of events beyond the SM predictions is observed in the $\ttbar$ invariant mass spectra, as shown in Fig.~\ref{fig:Plot1}(a). A heavy $Z^{'}$ boson of width 1\% is excluded for masses $m_{Z^{'}}$ < 3.0 TeV. The results are interpreted as well using simplified models of DM.

\medskip

A similar search performed by the CMS Collaboration uses 35.9~fb$^{-1}$ of $\sqrt{s}$ = 13 TeV $pp$ collision data~\cite{bib:CMS_res}. It considers all three decay modes of the $\ttbar$-system (single lepton, dilepton and fully hadronic) and uses reconstruction techniques that are optimised for top quarks with high Lorentz boosts, which requires the use of non-isolated leptons and jet substructure techniques. Events in the fully hadronic channel are required to have two high-$\pt$ top-tagged large-$R$ jets, and are separated into signal regions (SRs) based on the rapidity difference between the two jets, and the number of jets with a $b$-tagged subjet\footnote{A subjet is defined as a small-$R$ jet reconstructed within a larger-$R$ jet.}. Events in the single-lepton analysis are required to contain one high-$\pt$ nonisolated electron or muon and at least 2 high-$\pt$ small-$R$ jets; a boosted decision tree is used to suppress the main $W$+jets background. Events in the dilepton channel are categorised into boosted and non-boosted SRs and background-enriched region based on the sum of $\Delta R$ between the leading and subleading lepton and the nearest jet. No evidence for a massive $\ttbar$ resonance is found, as shown in Fig.~\ref{fig:Plot1}(b), and the analysis excludes narrow $Z^{'}$ bosons with masses up to 3.8 TeV. The contributions from the single-lepton and fully hadronic channels dominate the sensitivity over most of the mass range, apart from the region of lowest masses, where the dilepton channel makes a significant contribution, as shown in Fig.~\ref{fig:Plot1}(c).

\section{Searches for vector-boson resonances decaying to $t\bar{b}$-quarks}
\begin{figure*}[hb]
\centering
\begin{overpic}[height=0.27\textwidth]{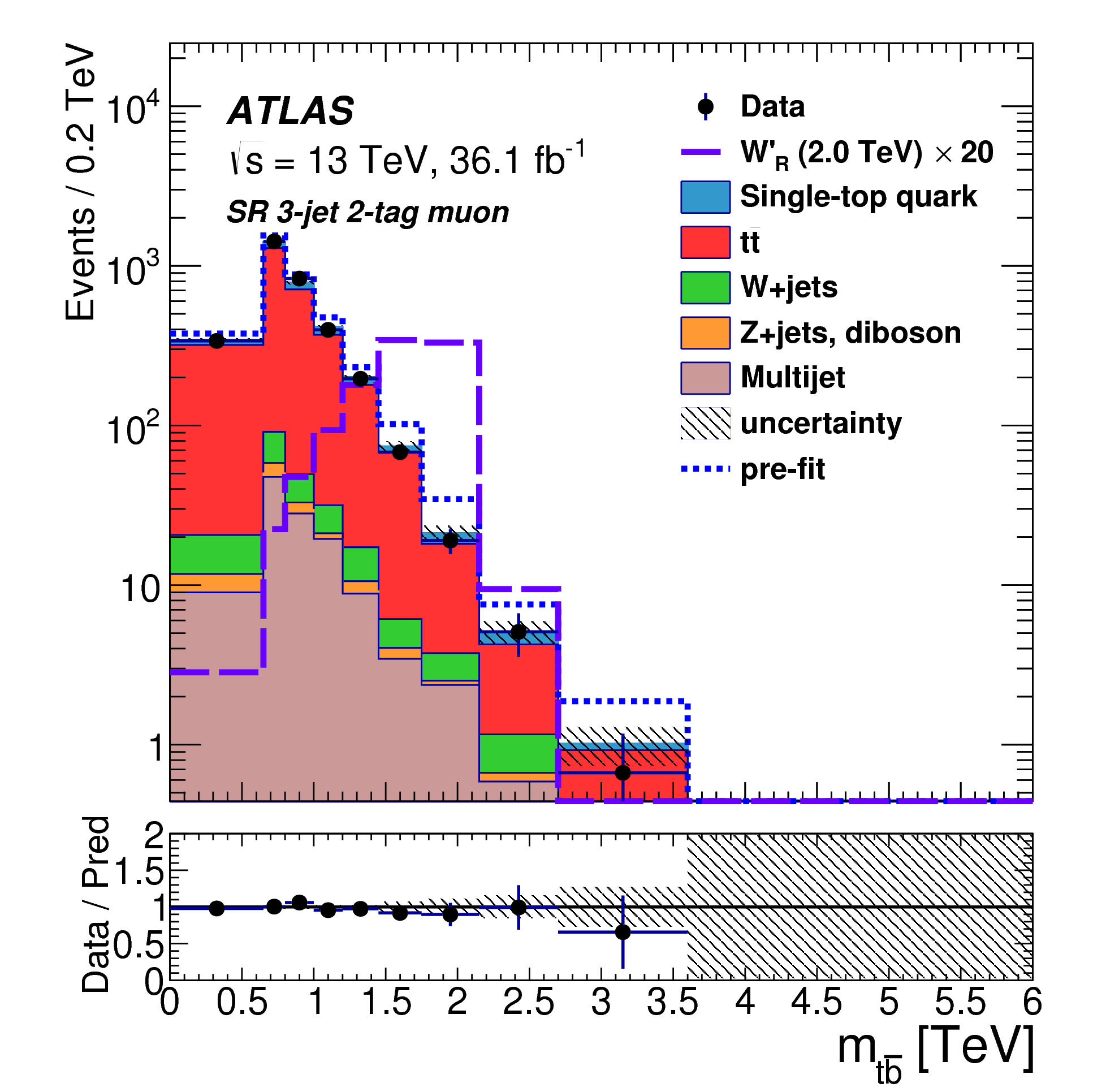}
\put(75,-7){(a)}
\end{overpic}
\qquad
\begin{overpic}[height=0.27\textwidth]{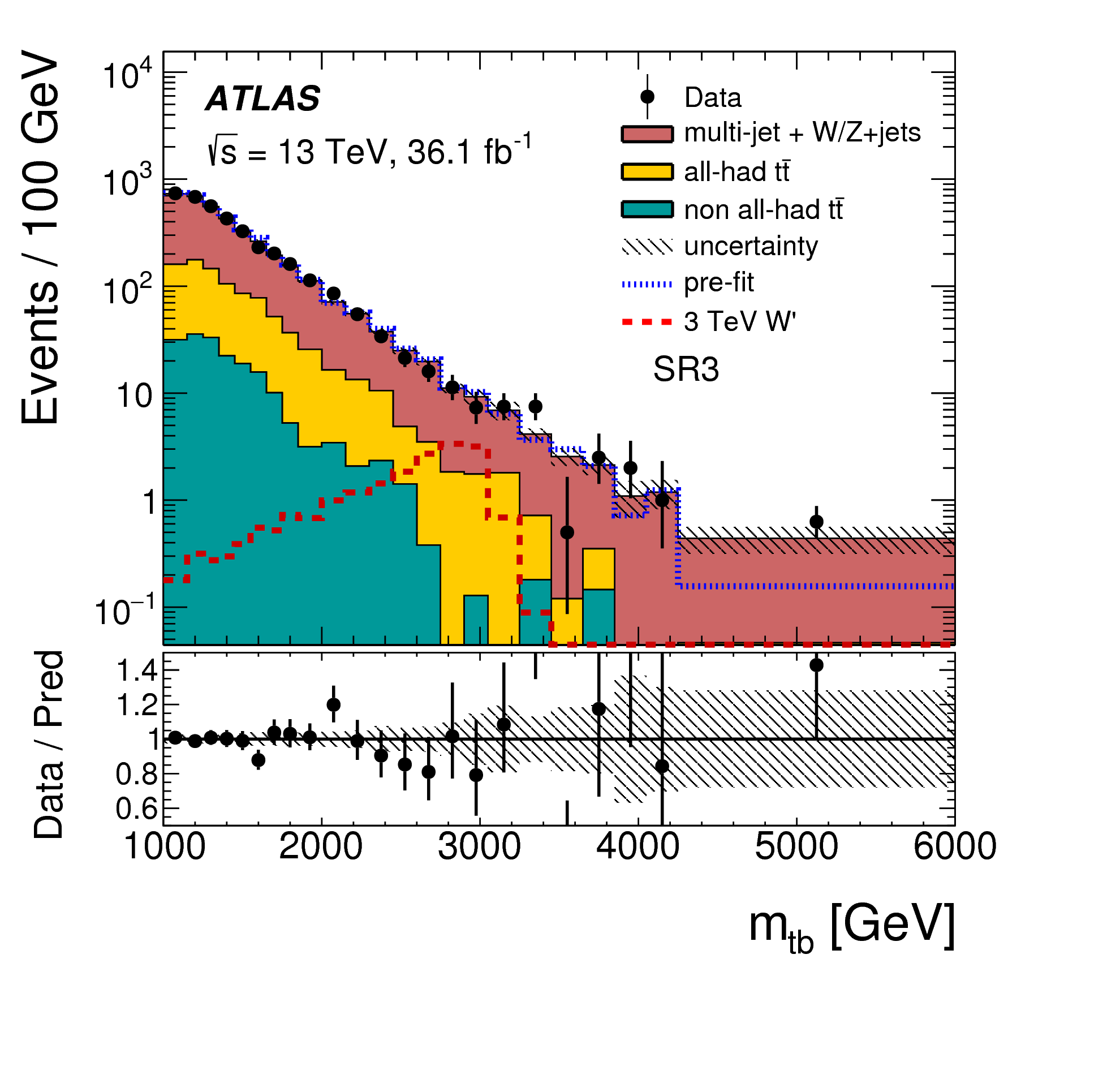}
\put(75,-7){(b)}
\end{overpic}
\qquad
\begin{overpic}[height=0.27\textwidth]{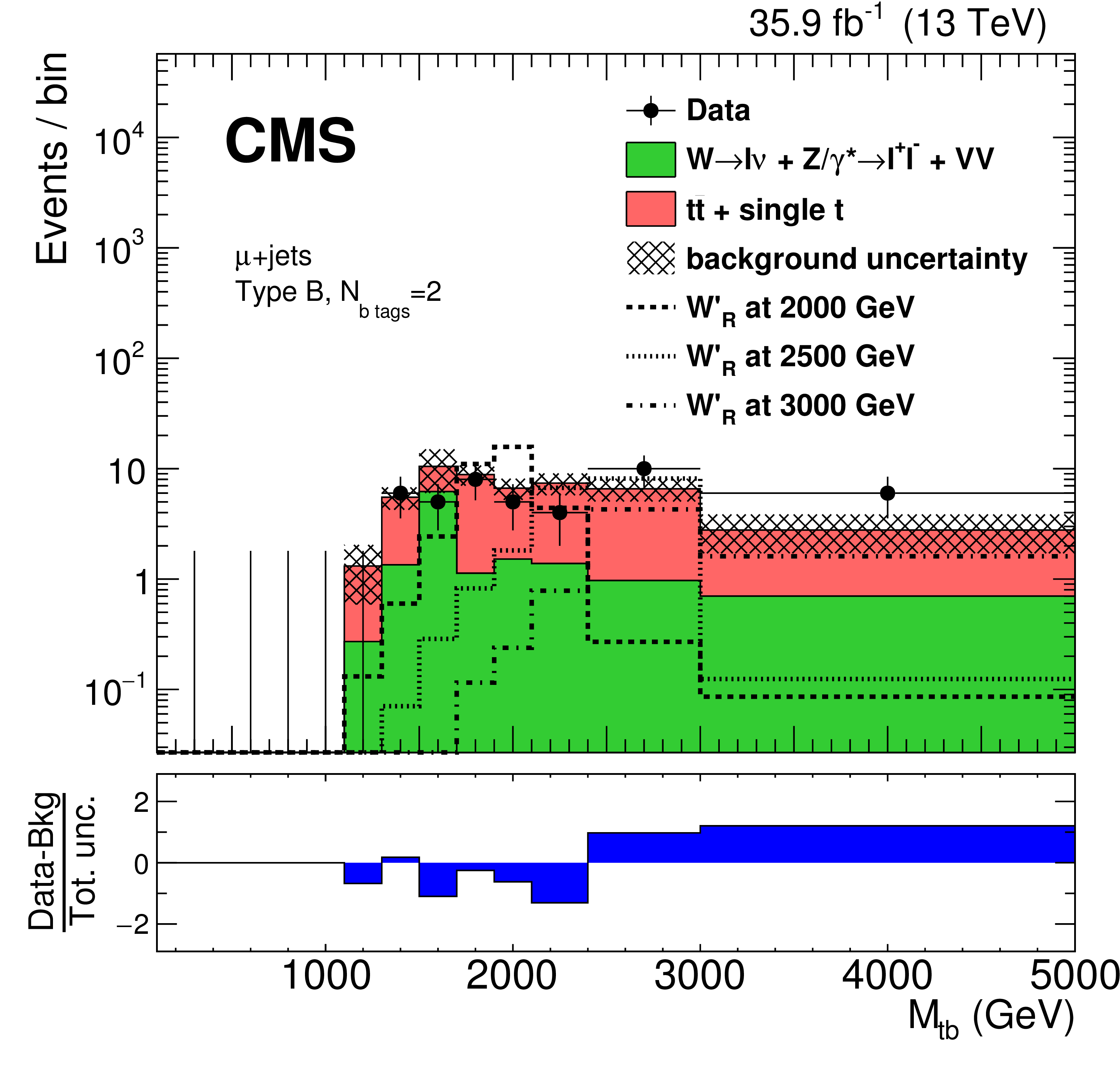}
\put(75,-7){(c)}
\end{overpic}
\caption{(a) Comparisons between data and prediction for $m_{t\bar{b}}$ in the (a) semi-leptonic~\cite{bib:ATLAS_Wprime_1L} and (b) fully hadronic~\cite{bib:ATLAS_Wprime_allhad} ATLAS searches, and (c) single-muon region of the CMS search~\cite{bib:CMS_Wprime}. Expected signal contribution corresponding to different $W^{'}$ boson masses is shown.
}
\label{fig:Plot2} 
\end{figure*}

A search for a $W^{'}$ boson decaying into a top and a bottom quark was performed using 36.1~fb$^{-1}$ of $\sqrt{s}$ = 13 TeV $pp$ collision data collected with the ATLAS experiment~\cite{bib:ATLAS_Wprime_1L}. The analysis focuses on the semi-leptonic decay channel, and candidate events are required to have exactly one charged lepton, two to four jets with at least one of them $b$-tagged and large missing transverse momentum ($\met$). The phase space is then subdivided into a SR and several validation regions enriched with $W$+jets, $\ttbar$ and $W$+heavy-flavour jets backgrounds. No significant excess over the background prediction is observed in the invariant mass spectrum of the top quark and bottom quark ($m_{t\bar{b}}$, as shown in Fig.~\ref{fig:Plot2}(a). For right-handed $W^{'}$ bosons with coupling to the SM particles equal to the SM weak coupling constant, masses below 3.15 TeV are excluded at the 95\% confidence level (CL). 

\medskip

The ATLAS Collaboration has also searched for $W^{'} \to t\bar{b}$ in the fully hadronic final state~\cite{bib:ATLAS_Wprime_allhad} using 36.1~fb$^{-1}$, using the same data sample as the analysis presented above. The fully hadronic search features a background dominated by QCD multijet production, which is estimated via data-driven methods. The analysis applies jet substructure optimised to select large-$R$ jets originating from hadronically decaying top quarks using the shower deconstruction algorithm and $b$-tagging of small-$R$ jets. The observed $m_{t\bar{b}}$ distribution is consistent with the background-only prediction, as shown in Fig.~\ref{fig:Plot2}(b).  As the two searches by the ATLAS experiment are complementary and use mutually orthogonal event selections, the results are combined and right-handed $W^{'}$ bosons with masses below 3.25 TeV are excluded for the combined semi-leptonic and hadronic scenarios~\cite{bib:ATLAS_Wprime_1L}.

\medskip

The CMS Collaboration reported results using a $\sqrt{s}$ = 13 TeV $pp$ data set of 35.9~fb$^{-1}$ searching for $W^{'} \to t\bar{b}$ in the semi-leptonic decay channel~\cite{bib:CMS_Wprime}. Events with exactly one high-$\pt$ nonisolated electron or muon, significant $\met$ and multiple jets in the final state are selected. Event categorisation based on the top quark candidate $\pt$ and the $\pt$ of the four-vector sum of the two leading $\pt$ jets improves the sensitivity to high signal masses. No significant excess over the background prediction is observed, as shown in Fig.~\ref{fig:Plot2}(c), and 95\% CL upper limit of 3.6 TeV on the mass of right-handed $W^{'}$ bosons is set.

\section{Searches for vector-like quarks}
\begin{figure*}[hb]
\centering
\begin{overpic}[height=0.3\textwidth]{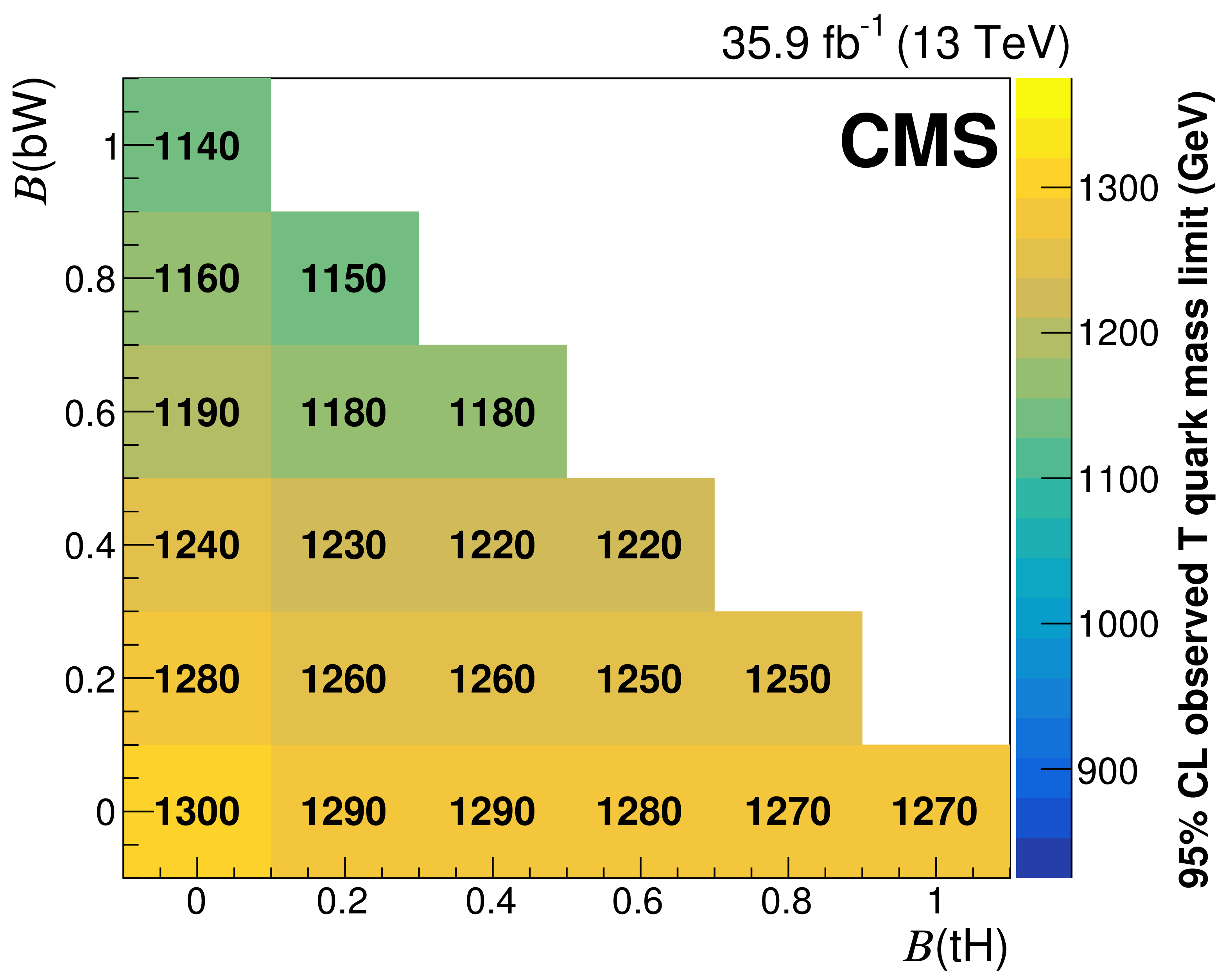}
\put(75,-7){(a)}
\end{overpic}
\qquad
\qquad
\begin{overpic}[height=0.3\textwidth]{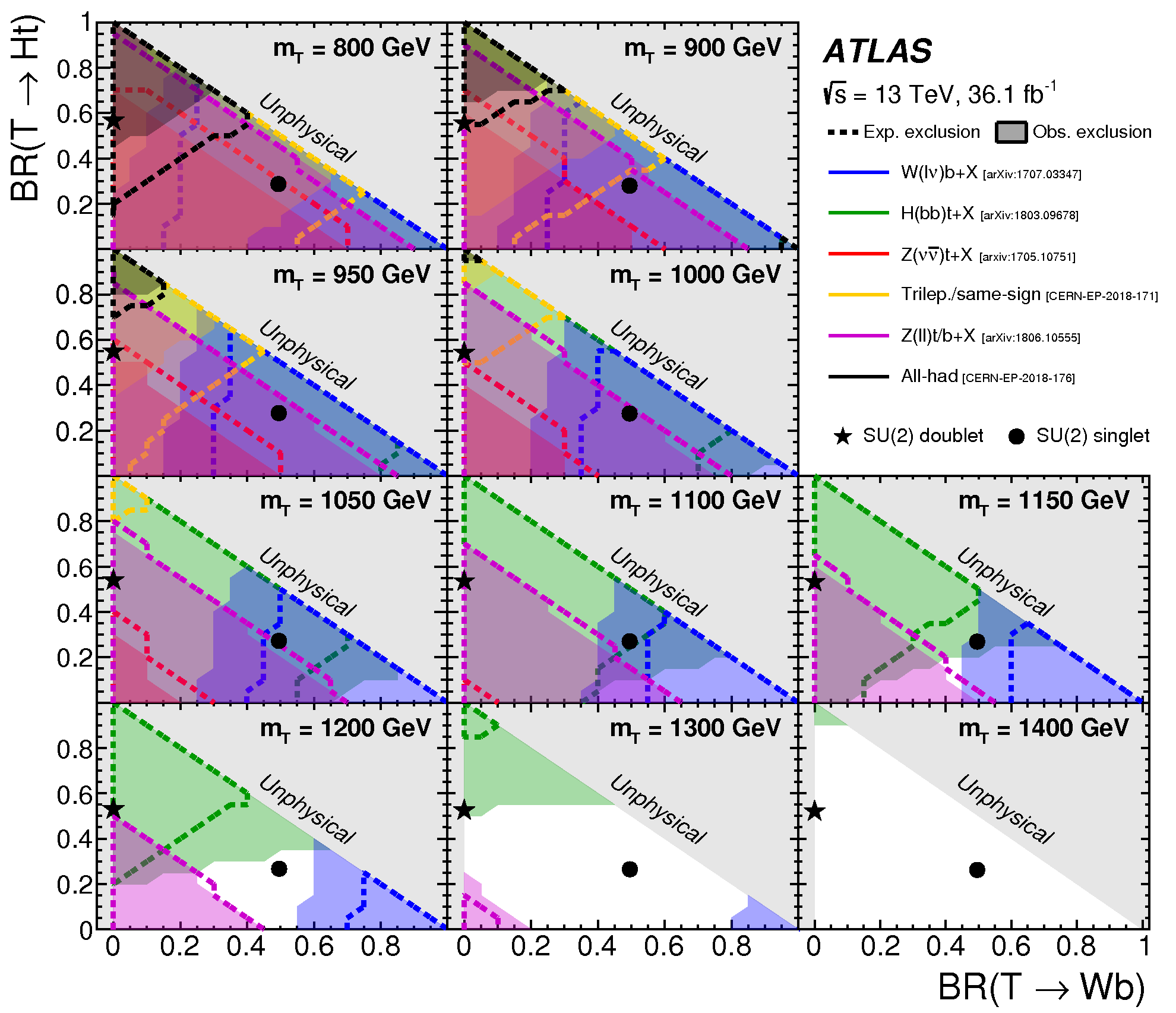}
\put(75,-7){(b)}
\end{overpic}
\caption{
(a) The 95\% CL observed lower limits on the $T$ quark mass, expressed in GeV, after combining all the CMS search channels in the plane of BR($T \to Ht$) versus BR($T \to Wb$)~\cite{bib:CMS_vlq}. (b) Observed (filled area) and expected (dashed line) 95\% CL exclusion in the plane of BR($T \to Ht$) versus BR($T \to Wb$), for different values of $T$ quark mass for the various analyses contributing to the ATLAS $T\bar{T}$ combination~\cite{bib:ATLAS_vlq_combo}.
}
\label{fig:Plot3} 
\end{figure*}

Recently, a generic search for pair production of heavy vector-like $T$ and $B$ quarks using a $\sqrt{s}$ = 13 TeV $pp$ data sample of 35.9~fb$^{-1}$ collected with the CMS experiment was performed~\cite{bib:CMS_vlq}. Three channels are considered, corresponding to final states with one lepton, two same-sign leptons, or at least three leptons. The analysis makes use of techniques to identify Lorentz-boosted hadronically decaying $W$ and Higgs bosons by quantifying the consistency of the jet’s internal structure with an N-prong hypothesis. The single-lepton channel has the highest efficiency for decay modes with at least one $T \to bW$ decay, the same-sign dilepton channel is sensitive to $B \to tW$ decays, and the trilepton channel has high efficiency for decay modes with at least one $T \to tZ$ decay. A MC simulation-based (data-driven) estimation of prompt (instrumental) backgrounds is used. A simultaneous fit in 16 single-lepton SRs, 6 single lepton control regions, event yields for the same-sign dilepton channel, and 4 trilepton categories is performed. Combining these channels, $T$~($B$) quarks at 95\% CL are excluded with masses below 1200 (1170) GeV in the singlet branching fraction (BR) scenario and 1280 (940) GeV in the doublet BR scenario, shown in Fig.~\ref{fig:Plot3}(a).

\medskip

The ATLAS Collaboration performed a combination of seven analyses searching for pair-produced vector-like partners of the top and bottom quarks in various decay channels ($T \to Zt/ Wb /Ht$, $B \to Zb / Wt / Hb$) using 36.1~fb$^{-1}$ of $pp$ collision data at $\sqrt{s}$ = 13 TeV~\cite{bib:ATLAS_vlq_combo}. The analyses contributing to the combination are: ``$H(bb)t+X$'' that targets $\ttbar$ events with at least one VLQ decaying into $Ht$, with $H \to b\bar{b}$; ``$W(l\nu)b+X$'' that targets $T\bar{T} \to Wb Wb$ events with one $W$-boson decaying leptonically and the other hadronically; ``$W(l\nu)t+X$'' which is optimised to target $B\bar{B}$ signals, especially in the case where $B \to Wt$; ``$Z(\nu \nu)t+X$'' that targets $T\bar{T} \to Zt Zt $ events with an invisible Z decay; ``$Z(ll)t/b+X$'' search for $T\bar{T}$ and $B\bar{B}$ events containing a leptonically decaying $Z$ boson; ``trilep./same-sign'' that targets $T\bar{T}$ and $B\bar{B}$ decays with multilepton final states; ``Fully hadronic'' search that focuses on final states with zero leptons, with significant sensitivity to $B\bar{B} \to Hb H\bar{b}$. Upper limits are set on the production cross-sections for $T\bar{T}$ and $B\bar{B}$ as a function of the VLQ mass. Singlet T (B) quarks at 95\% CL excluded for masses below 1.31 (1.22) TeV and $m_{T(B)}$ < 1.37 (1.37) TeV in the doublet scenario, shown in Fig.~\ref{fig:Plot3}(b).

\section{Search for dark matter produced in association with a $\ttbar$-pair}
\begin{figure*}[ht]
\centering
\begin{overpic}[height=0.27\textwidth]{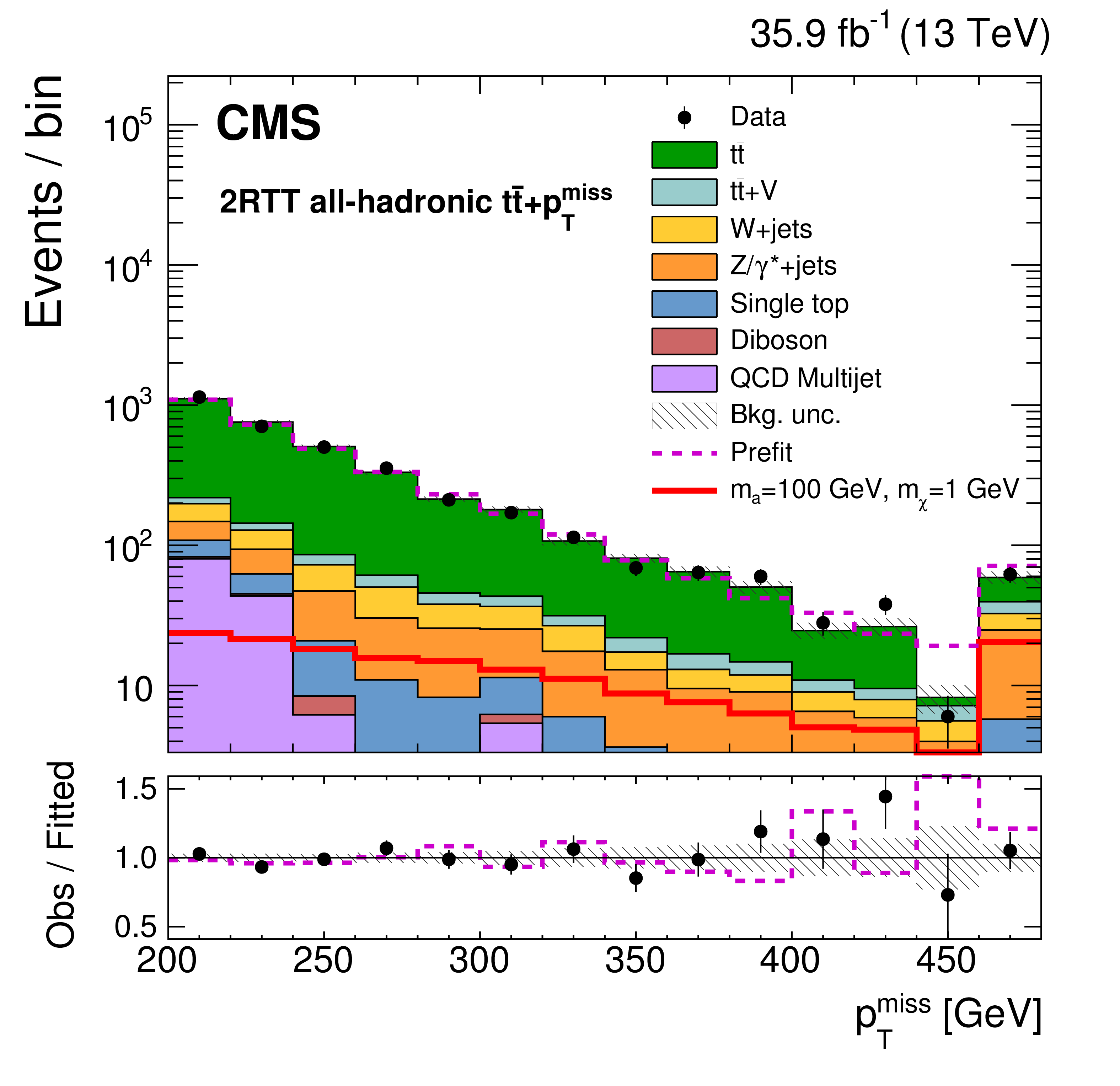}
\put(75,-7){(a)}
\end{overpic}
\qquad
\begin{overpic}[height=0.27\textwidth]{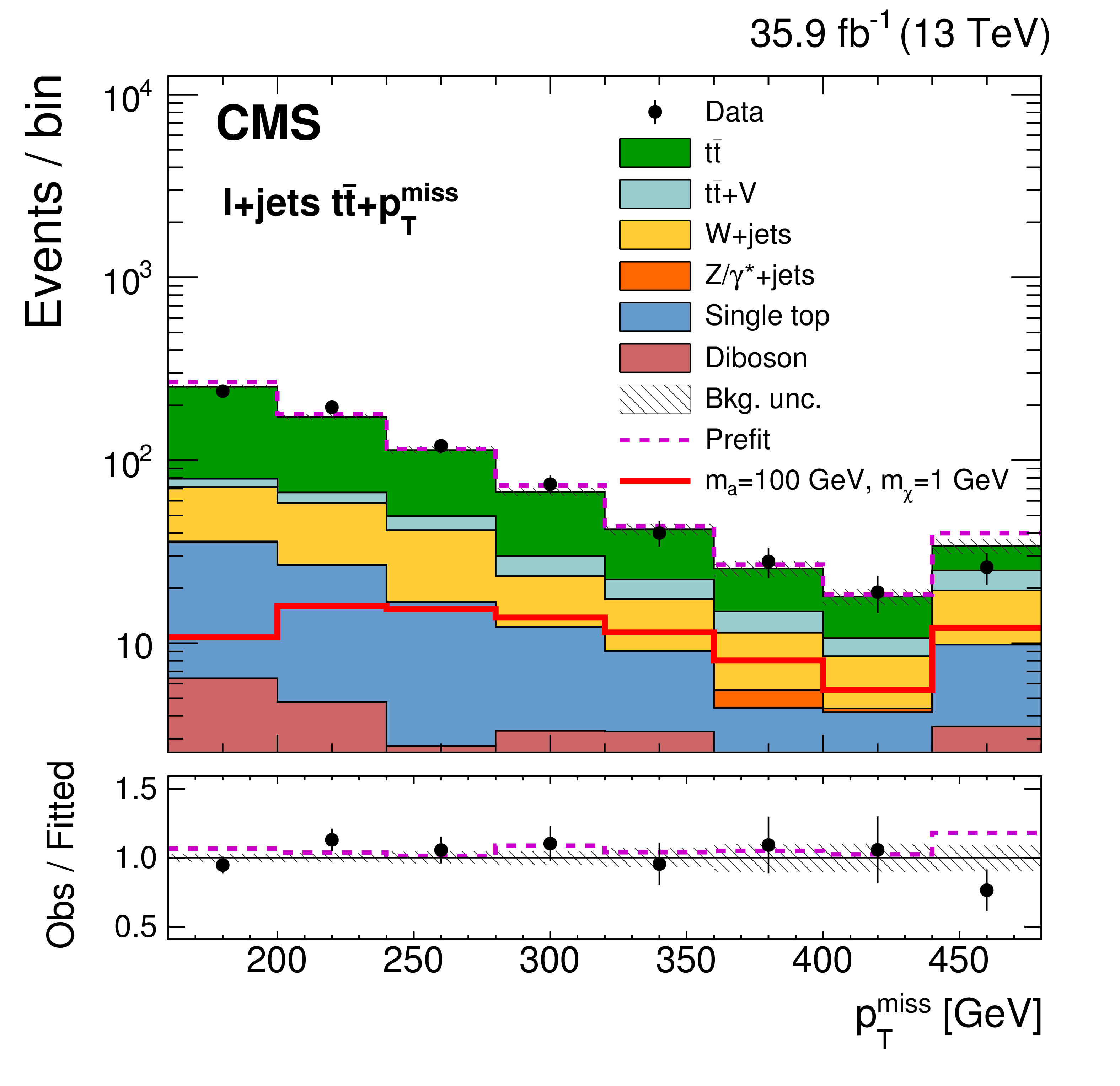}
\put(75,-7){(b)}
\end{overpic}
\qquad
\begin{overpic}[height=0.27\textwidth]{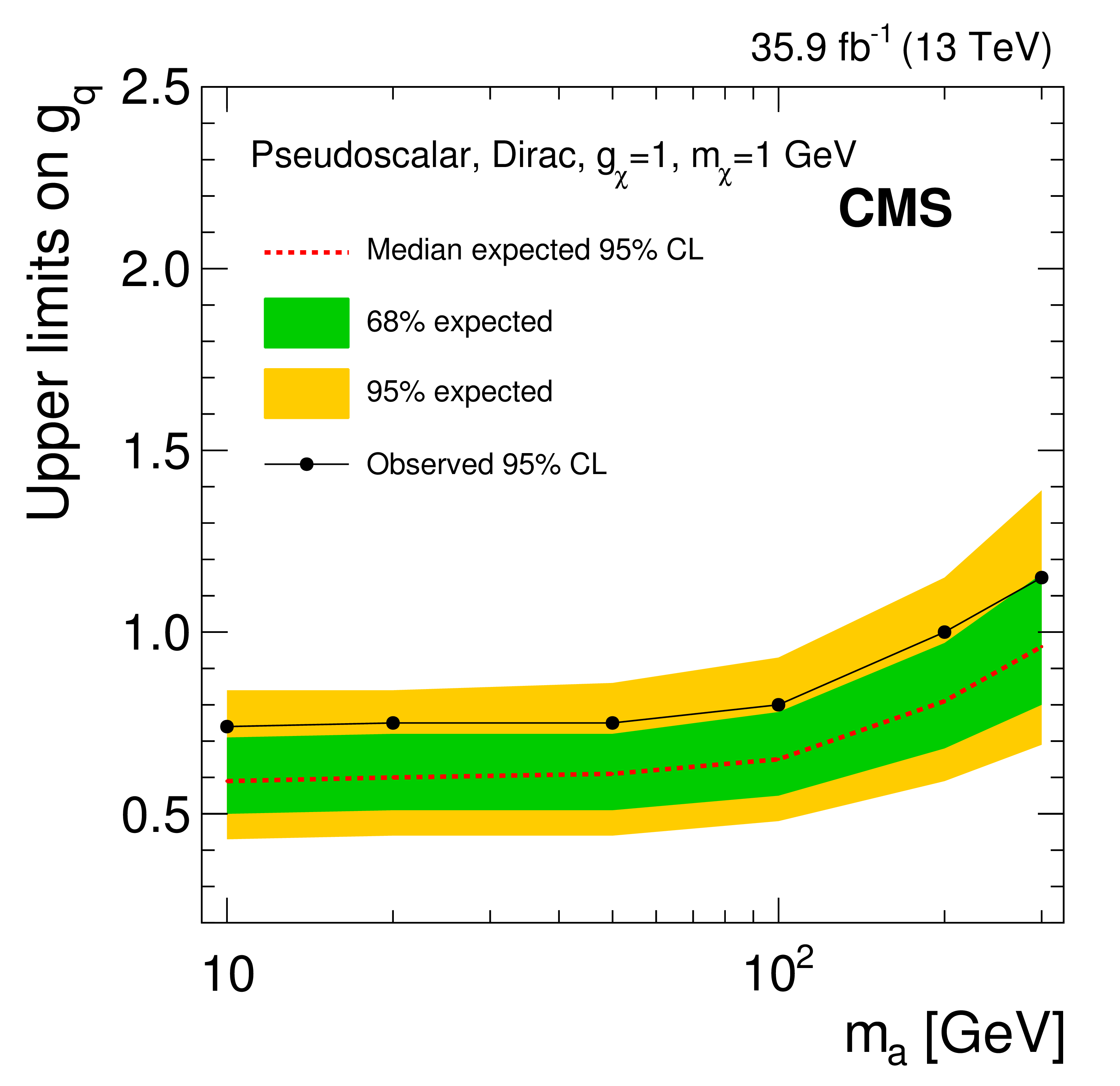}
\put(75,-7){(c)}
\end{overpic}
\caption{Comparison between data and prediction for the $p_{\rm{T}}^{\rm{miss}}$ distribution in the (a) all-hadronic and (b)~lepton+jets SRs. The solid red line shows the expectation for a signal with $m_{a}$ = 100 GeV and $m_{\chi} = 1$~GeV. (c) 95\% observed and median expected CL upper limits on the coupling strength of the pseudoscalar mediator to the SM quarks under the assumption that $g_{\chi} = 1$ and $m_{\chi} = 1$~GeV~\cite{bib:CMS_dm}.}
\label{fig:Plot4} 
\end{figure*}

A search for $\ttDM$ signal, which results in high-$\pt$ jets, leptons, and significant $\ptmiss$, was recently performed by the CMS Collaboration using 35.9~fb$^{-1}$ of $\sqrt{s}$ = 13 TeV $pp$ collision data~\cite{bib:CMS_dm}. The search covers all three $\ttbar$ decay modes. In the all-hadronic channel, a multivariate discriminant based on jet properties and kinematic information is used to identify top quarks that decay into three jets. The multijet background is suppressed by requiring several kinematical cuts. In the lepton+jets channel, control regions enriched in both $Z(\nu\bar{\nu})$+jets and $W(\ell\nu)$+jets contributions are used to improve the simulation-based background estimates. In the dilepton channel, separate categories are considered for events with same- and different-flavour lepton pairs. The estimates for the backgrounds from Drell--Yan production and from jets misidentified as leptons are performed using dedicated sideband regions not included in the fit. Using several selection requirements, two all-hadronic, one lepton+jets and four dilepton signal regions are defined. All three channels are used in a simultaneous maximum-likelihood fit of $\ptmiss$ distributions to extract a potential DM signal. No significant excess over the SM expectation is observed, as can be seen in Figs.~\ref{fig:Plot4}(a, b). The observed (expected) upper limits exclude scalar and pseudoscalar masses of 160 (240) and 220 (320) GeV, respectively, at 95\% CL under the $g_{q}=g_{\chi}=1$ benchmark scenario, as shown in Fig.~\ref{fig:Plot4}(c).

\section{Summary}
Both ATLAS and CMS Collaborations have a broad search program for BSM physics using top/$b$-quarks, and challenging techniques and rare decay modes are being exploited. Searches are using all possible $\ttbar$ final states, with dedicated resolved and boosted channels. In the latter, jet substructure and reclustering techniques are being used. Latest multivariate methods are used for background rejection, and combination of results of various complementary searches in order to set even strongest limits on new physics models are performed. No evidence for new physics has emerged yet, and we all are looking forward to new results using the full Run 2 data set, thanks to the great performance of the LHC.

\end{document}